\documentclass[12pt,dvips]{article}
\usepackage{epsfig}

\begin{document}

\title{Reality or Locality? - Proposed test
to decide \textit{how} Nature breaks Bell's inequality}

\date{}

\author{Johan Hansson\footnote{c.johan.hansson@ltu.se} \\
Department of Physics \\ Lule{\aa} University of Technology
\\ SE-971 87 Lule{\aa}, Sweden}

\maketitle

\begin{abstract}
Bell's theorem, and its experimental tests, has shown that the two
premises for Bell's inequality - locality and objective reality -
cannot both hold in nature, as Bell's inequality is broken. A
simple test is proposed, which for the first time may decide which
alternative nature actually prefers on the fundamental, quantum
level. If each microscopic event is truly random (\textit{e.g.} as
assumed in orthodox quantum mechanics) objective reality is not
valid, whereas if each event is described by an unknown but
deterministic mechanism (``hidden variables") locality is not
valid. This may be analyzed and decided by the well-known
reconstruction method of Ruelle and Takens; in the former case no
structure should be discerned, in the latter a reconstructed
structure should be visible. This could in principle be tested by
comparing individual ``hits" in a double slit experiment, but in
practice a single fluorescent atom, and its (seemingly random)
temporal switching between active/inactive states would possibly
be better/more practical, easier to set up, observe and analyze.
However, only imagination limits the list of possible experimental
setups.
%to test the possibility of a deterministic, but chaotic, origin
%of quantum mechanical randomness.
%\\
%PACS numbers: 03.65.-w, 03.65.Bz, 03.65.Ud
\end{abstract}

\begin{center}
 \end{center}

%\pacs{ PACS numbers: 03.65.-w, 03.65.Bz, 03.65.Ud}
\newpage

Through Bell's theorem \cite{Bell},\cite{CHSH}, which put the
(in)famous Einstein-Podolsky-Rosen \cite{EPR} argument on a solid
and testable footing, and experimental tests thereof
\cite{Clauser},\cite{Aspect1},\cite{Aspect2},\cite{Tittel},\cite{Zeilinger}
it has been proven beyond reasonable doubt that no ``locally
realistic" fundamental model of the world can be correct. That is,
a ``sensible" world-view, such as that proposed in \cite{EPR}, is
unfortunately untenable.

So \textit{either} the objective reality-condition (that things
exist in definite states whether we look or not) must be broken,
\textit{e.g.} as in orthodox quantum mechanics, \textit{or} the
locality-condition (that events arbitrarily far away cannot affect
what happens here and now - relativistic separability and
causality) must be broken, \textit{e.g.} as in non-local hidden
variable theories. The variables are called ``hidden" because
their existence is only conjectured and beyond our (present)
control, but meant to complete quantum mechanics into a uniform
description of micro and macro\footnote{Bell himself was heavily
biased towards a hidden variable resolution of the problem
\cite{Unspeakable}.}. The first detailed such theory, perfectly
deterministic and compatible with all known experimental data, was
\cite{Bohm}. Notice, however, that we are not necessarily
considering any specific existing hidden variable theory, but an
``ultimate" hidden variable theory that \textit{in principle}
decides everything deterministically. In contrast in the orthodox
approach to quantum mechanics the quantum particles in effect
behave as particles when observed and as waves when not observed -
thereby, and at the most fundamental level, introducing the
ill-defined act of observation (``measurement problem", ``collapse
of the wave function", Bohr's ``irreversible act of measurement"),
whereas particles in hidden variable theories always behave as
particles but are being ``pushed around" by the underlying
(hidden) dynamics. In such deterministic systems the present state
completely and uniquely determines the future, but as is
well-known chaotic systems can ``impersonate" randomness due to
their extreme sensitivity to initial conditions; in a nutshell
chaos is about order and disorder in deterministic systems that
are nonlinear.

So far, it has not been possible to distinguish between the
locality vs. reality alternatives, and the choice has been mainly
one of personal taste.

However, as hidden variable theories are deterministic (quantum
particles behaving as realistic classical particles all the time,
encoding Einstein's ``elements of physical reality" \cite{EPR})
and orthodox quantum mechanics fundamentally probabilistic (each
individual event/measurement assumed to be completely random), it
should be possible to experimentally test the distinction between
them.

An experiment to test this possibility could be devised in analogy
to the confirmation of deterministic chaos in a dripping water
faucet \cite{Shaw},\cite{Shaw2}. It is of course well-known that
deterministic chaos requires nonlinear systems whereas the
Schr\"{o}dinger equation is linear. However, most hidden variable
theories like the original by Bohm \cite{Bohm} are manifestly
nonlinear\footnote{As an aside, if hidden variables is the correct
way to explain the violation of Bell's inequality this could make
\textit{true} quantum chaos possible, as opposed to the usual
notion of ``quantum chaos" which is concerned with quantum
signatures of corresponding systems known to be chaotic in the
classical case, as the linear structure of the Schr\"{o}dinger
equation alone does not support true chaos.}.

If we, for example, replace the dripping faucet with a double-slit
experiment\footnote{According to R.P. Feynman the double slit
experiment ``...has in it the heart of quantum mechanics. In
reality, it contains the \textit{only} mystery.", The Feynman
Lectures on Physics, Vol.III, p. 1-1.} with individual quantum
entities (electrons, neutrons, photons, etc), the effectively
one-dimensional position ($q_i$) of the successive ``hits" on the
detector screen, in effect defining a discrete time-series, can be
used to try to reconstruct a chaotic attractor, in case the
underlying theory is dissipative
%\footnote{For the orthodox quantum mechanical
%``measurement problem" it seems advantageous if the mechanism is
%dissipative, as the phase space inevitably shrinks upon
%measurement, this is however irrelevant for the purpose of the
%present article.}
, or a deterministic structure in phase space, in case it is
non-dissipative (Hamiltonian), by applying a method
\cite{Packard},\cite{Takens} of converting a single data series
into a phase space portrait via ``delay coordinate embedding".
This can be accomplished, assuming a suitably low-dimensional
attractor/structure, by defining the coordinates as follows
\begin{equation}
x = q_i, \; \; \;
%\end{equation}
%\begin{equation}
y = q_{i+1}, \; \; \;
%\end{equation}\begin{equation}
z = q_{i+2}.
\end{equation}
A given $i$ then gives a point, $(x,y,z)$, in phase space.

To give an elementary example, the seemingly random data in Fig.1
is really due to the deceptively simple, but actually incredibly
rich, ``logistic mapping"
\begin{equation}
x_{n+1} = k \, x_n (1 - x_n),
\end{equation}
in its highly chaotic regime with $k = 4$ \cite{May}.

\begin{figure}[h]
\begin{center}
\psfig{file=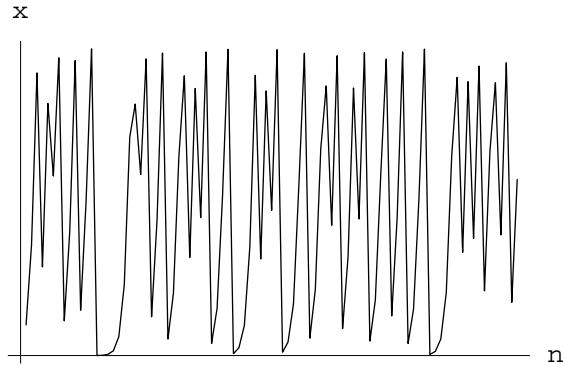}
%\leavevmode
%\epsffile{logistic.eps}
\end{center}
 \caption{Seemingly random data, actually generated by the
 simple and deterministic
 ``logistic mapping" in its chaotic region, see text, and \cite{May}.}
 \end{figure}

The reconstructed attractor, using the method described above, is
seen in Fig.2 (2-D) and in Fig.3 (3-D).

 \begin{figure}[h]
\begin{center}
\epsfig{file=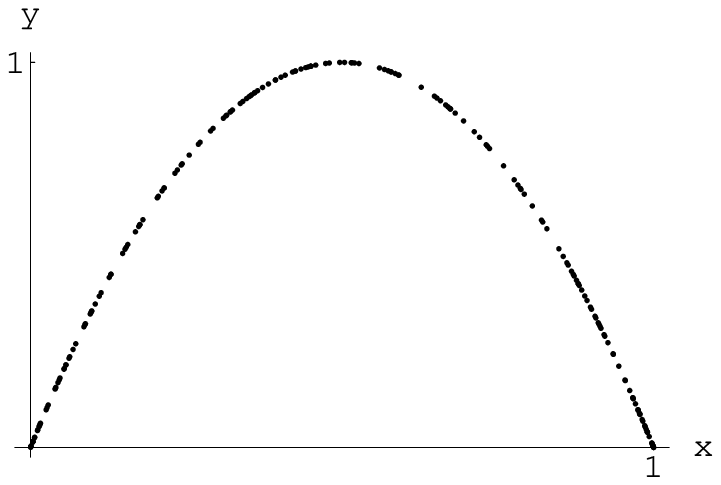}
%\epsffile{2Dlogistic.eps}
\end{center}
 \caption{The reconstructed attractor in 2-D from the data in
 Fig.1, using Eq. (1),
 showing that its ``randomness" has its origin in dynamical
 \textit{deterministic} chaos.}
 \end{figure}

  \begin{figure}[h]
\begin{center}
\epsfig{file=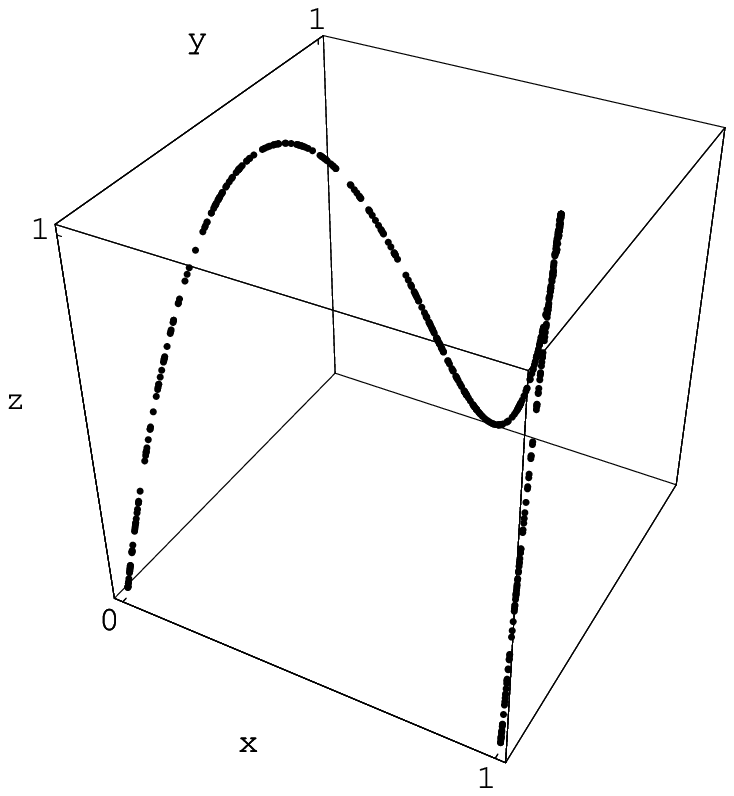}
\end{center}
 \caption{Reconstructed attractor in 3-D from the data in Fig.1, again using Eq. (1).}
 \end{figure}

We do not, however, expect that an eventual attractor/structure in
real quantum mechanical data will be so simple and
low-dimensional, even though the logistic mapping \textit{has}
been shown to be in qualitative \textit{and} quantitative
agreement with numerous real-life systems in all branches of
science, see \textit{e.g.} \cite{Cvitanovic} for some early
examples. This would be very surprising if not for the remarkable
fact that there exists a ``universality" in this kind of chaos
\cite{Feigenbaum}.

In a sense, the logistic mapping is just like a model for
observing ``random" hits on an effectively 1D-detector screen of
unit length (arbitrarily defined), just like in the double slit
experiment. The detector in effect defines a natural Poincar\'{e}
section - a discrete ``stroboscope" mapping of the unit interval
onto itself - of the underlying continuous dynamics described by
differential equations. If there is a deterministic mechanism
underlying the ``random" hits on the screen, creating the known
statistical distribution after \textit{many} hits, it should then
show up as a structure in reconstructed phase space.

 In principle, to capture all emitted quantum particles, the ideal would be to have a perfectly efficient
 4$\pi$-detector, faithfully recording each individual quantum
 particle on its ``latitude and longitude". A 2D-iterated mapping, of the classic
 predator-prey kind, would then be a model for the successive hits, the
 simplest one using ``non-overlapping generations", where each hit
 is described by two coordinates (originally the populations of predator and prey
 species) and is determined by the previous hit through a
 mapping of the form
 \begin{equation}
x_{n+1} = f(x_n, y_n), \; \; \;
%\end{equation}
%\begin{equation}
y_{n+1} = g(x_n, y_n).
%\end{equation}\begin{equation}
\end{equation}
One such model, the H\'{e}non mapping \cite{Henon}
 \begin{equation}
x_{n+1} = y_n + 1 - a x_n^2 , \; \; \;
%\end{equation}
%\begin{equation}
y_{n+1} = b x_n ,
%\end{equation}\begin{equation}
\end{equation}
gives the famous H\'{e}non-attractor. For the canonical values $a$
= 1.4 and $b$ = 0.3  the H\'{e}non map is chaotic; each individual
hit appears random, but a clear structure builds up over time,
analogous to hits in the double-slit experiment. In the former
case the structure is fractal \cite{Mandelbrot}, whereas in the
latter case it may or may not be.

However, one could argue that any eventual hidden variables must
``know" that we have restricted the ``landing platform" for the
quantum particle to an effectively 1D-strip, so that additional
spatial variables are superfluous. The hidden variables must also
keep track of if one or both slits are open and relay that
information non-locally (faster than the speed of light) to the
detector screen, as in \cite{Bohm}, to comply with the violation
of Bell's inequality.

As modern technology has made it possible to trap and observe
individual quantum objects, such as atoms, it might be better and
easier to exploit this fact than trying to use the mythical
double-slit. Measurements of ``quantum jumps" in single atoms
\cite{Nagourney},\cite{Bergquist}, and the resulting fluctuation
of their fluorescent on/off-states, may make an ideal testing
ground where recorded data should already be present (the
time-series underlying Fig. 2 in both articles
\cite{Nagourney},\cite{Bergquist} could in principle be directly
inserted into Eq. (1) above), but only imagination limits the list
of possible experimental setups.

\textit{If} the seemingly random florescence gives rise to a
distinct structure in phase space, with non-integer fractal
dimension, onto which the phase space points are concentrated, it
would be a clear indication that it is actually the consequence of
dynamical deterministic chaos (\textit{i.e.} hidden variables), in
direct analogy to how \cite{Shaw},\cite{Shaw2} revealed
deterministic chaos in the dynamics of the dripping water faucet.
For examples of qualitatively typical chaotic
attractors/structures see, \textit{e.g.}, the figures in
\cite{Shaw},\cite{Shaw2} or the famous examples presented in the
figures in this article (accompanied by their respective physical
implications to the problem at hand in the figure captions to
Figs. 5 and 7). However, the exact \textit{shape}, dimension and
complexity will be governed by the (unknown) detailed underlying
dynamics. The rest of the analysis carries through just like in
\cite{Shaw},\cite{Shaw2}.

In fact, in the present case it is in principle even easier to
obtain a conclusive result as \textit{any} observed structure
indicates a deviation from the usual assumption of total
randomness of quantum mechanics - where it is normally assumed
that, \textit{e.g.}, the hit of an individual particle is a
completely independent and truly random process - even if one has
collected one million successive data points the next one,
according to orthodox quantum mechanics, will be a complete
surprise and impossible to predict even in principle, see Fig.4.

\begin{figure}[h]
\begin{center}
\epsfig{file=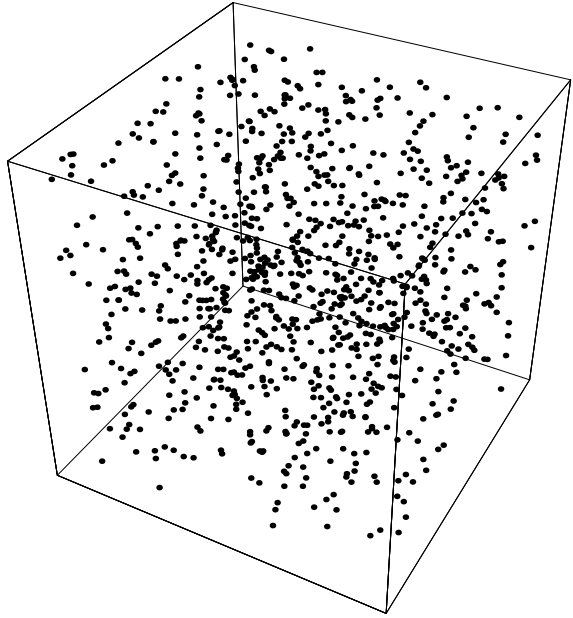}
\end{center}
 \caption{When no dynamical relation between the data points ($q_i$) exists,
 \textit{no} structure is obtained by the reconstruction mechanism, Eq. (1).
 This would be the case for ``orthodox" quantum mechanics where each individual hit/result/event is assumed to be completely random.
 The world could then \textit{not} be objectively real, but could be local.}
 \end{figure}

So, in a perfect world it should be easy to potentially disprove
orthodox quantum mechanics. A practical problem is of course that
there exist no perfect particle detectors, which results in
missing part of the series and also in the introduction of noise
in the data. The more of the series one misses, the harder it
becomes to reconstruct an (eventual) attractor/structure. This
may, as stated above, be circumvented by observing, \textit{e.g.},
single atoms exhibiting quantum jumps as this ``...can be detected
with unity quantum efficiency" \cite{Nagourney}.

If, however, no attractor/structure is found in the experimental
data, \textit{i.e.}, if the points are scattered randomly in phase
space, as in Fig.4, where every $q_i$ has been generated at
random, then quantum mechanical ``measurements" (\textit{e.g.},
hits on detector screen, timing between on/off-states, etc)
probably \textit{cannot} be described by deterministic equations,
and some truly stochastic effect(s) must instead be at work,
\textit{e.g.} as assumed in orthodox quantum mechanics.

\begin{figure}[h]
\begin{center}
\epsfig{file=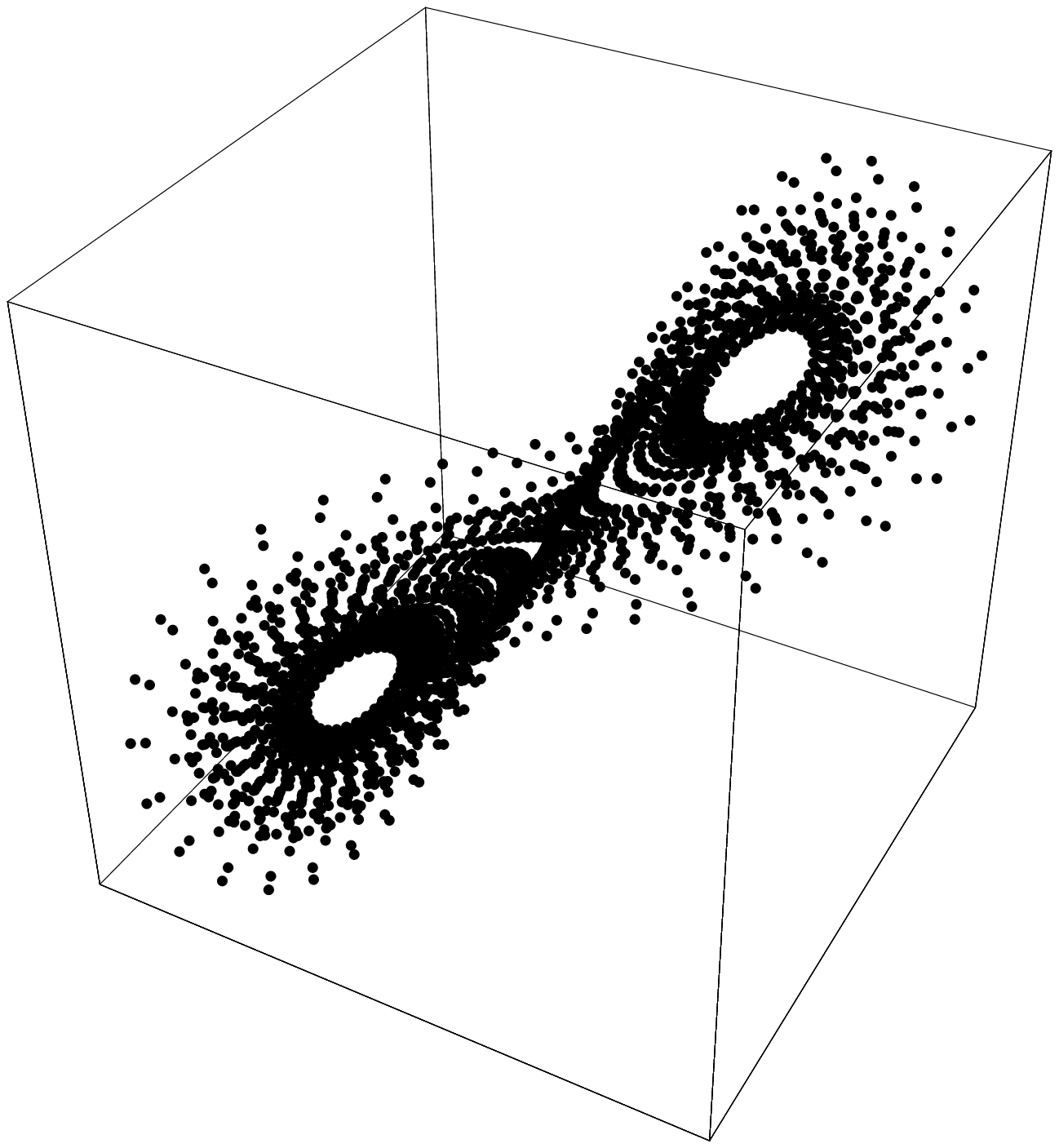, scale = 0.6}
\end{center}
 \caption{If ``hidden variables" are governed and determined by dissipative equations an attractor will be reconstructed by the
 $q_i$s. To mimic the apparently random behavior of quantum
 mechanical data it will be a ``strange attractor" with non-integer (fractal \cite{Mandelbrot}) dimension analogous to
 the famous Lorenz-attractor, here \textit{reconstructed} from time series data from only \textit{one} of the three variables of the Lorenz system \cite{Lorenz} - the very first concrete example of
 dissipative chaos. Any apparent attractor structure would tell us Nature is \textit{not} local - causes
 arbitrarily far may affect results ``here" - \textit{i.e.} there \textit{are} influences going faster than light (even if we cannot control them for practical telegraphy).
 Furthermore, it would indicate that the orthodox (``Copenhagen") interpretation of quantum mechanics is \textit{wrong}.}
 \end{figure}

 \begin{figure}[h]
\begin{center}
\epsfig{file=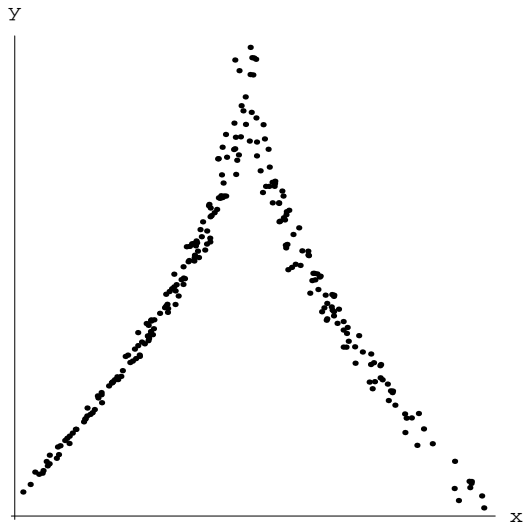, scale = 0.5}
\end{center}
 \caption{``The Lorenz Map" - when successive, erratically fluctuating, amplitude maxima were plotted for the Lorenz attractor (previous figure)
 using a technique analogous to the one described in this article, the surprising result was this nearly one-dimensional attractor; hidden order in chaos \cite{Lorenz},
 and a concrete simple example of the relation between continuous dynamics and discrete mappings.}
 \end{figure}

 \begin{figure}[h]
\begin{center}
\epsfig{file=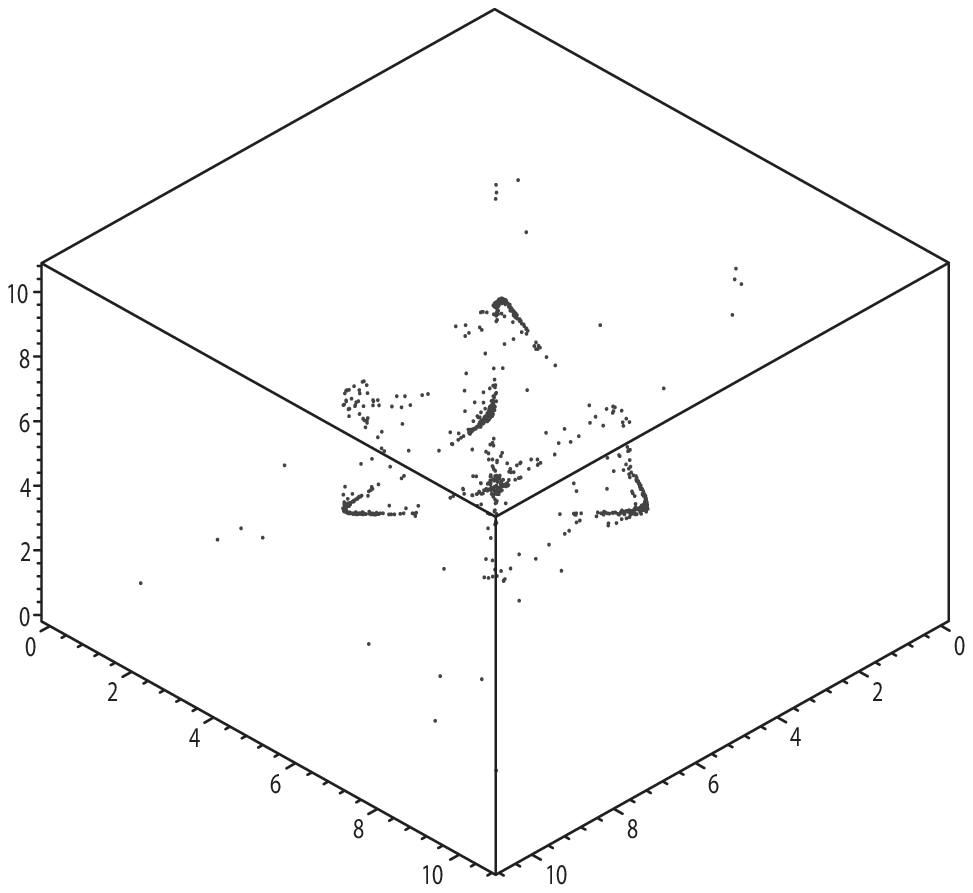, scale = 0.725, angle=0}
\end{center}
 \caption{If ``hidden variables" are governed and determined by non-dissipative (Hamiltonian) equations no attractor will result,
 yet a structure differing from pure randomness will emerge. Such a result would imply the same conclusion regarding the world as noted in the text accompanying Fig.5.
 %Instead a chaotic (fractal) structure in, \textit{e.g.}, a Poincar\'{e} section -
 %a ``stroboscopic slice" turning the full continuous behavior in phase space into a discrete mapping between return points - will
 %be reconstructed by the
 %$q_i$s., again to mimic the apparently random behavior of quantum
 %mechanical data.
 The example shows the modern \textit{reconstructed} phase space of the ``restricted circular three-body problem" in astronomy \cite{Gidea}, where Poincar\'{e}
 first glimpsed what today is known as deterministic chaos, non-dissipative in this case.
 This (the corrected and printed version \cite{Poincare}) was his winning contribution (price money: 2,500 Swedish Kronor) to a contest announced in
 1885 to celebrate the 60th birthday of the Swedish King Oscar II in
 1889. What Poincar\'{e} found was that small changes in the initial conditions (such as positions and initial velocities of planets) produced huge and unpredictable outcomes -
 deterministic chaos in today's parlance.
 }
 \end{figure}

Hence, it should be possible to test, and potentially falsify:
\textit{either} the hypothesis that quantum randomness is due to
underlying deterministic dynamics - hidden variables (in which
case the ``randomness" actually would merely be apparent, not
fundamental) - without having to know and penetrate the details of
the underlying equations, \textit{or} the standard fundamentally
probabilistic interpretation/postulate of Born as used in orthodox
quantum mechanics, and hence answer if Nature prefers to break
locality or objective reality on her fundamental level. In case of
the former, it would indicate an unexpected and deep hidden
connection between the three great revolutions of 20th-century
science; relativity, quantum mechanics and chaos theory, and
perhaps even point the way towards a unified complex nonlinear
systems theory of the future.

\end{document}